\documentclass[12pt,thmsa]{article}
\usepackage{amsfonts}

\usepackage{sw20lart}

\input{tcilatex}
\begin{document}

\begin{titlepage}
\begin{center}
\vspace{1cm}
\hfill
\vbox{
    \halign{#\hfil         \cr
         hep-th/0206170 \cr
           IPM/P-2002/018\cr
           June, 2002\cr} }
      
\vskip 1cm
{\Large \bf
Generalized Noncommutative Supersymmetry from a New Gauge Symmetry  }
\vskip 0.5cm
{\bf Reza Abbaspur\footnote{e-mail:abbaspur@theory.ipm.ac.ir} 
}\\ 
\vskip .25in
{\em
Institute for Studies in Theoretical Physics and Mathematics,  \\
P.O. Box 19395-5531,  Tehran,  Iran.\\}
\end{center}
\vskip 0.5cm

\begin{abstract}
Using the notion of a gauge connection on a flat superspace, we construct a general
class of noncommutative ($D=2,$ $\mathcal{N}=1$) supertranslation
algebras generalizing the ordinary algebra by inclusion of
some new bosonic and fermionic operators. We interpret the new operators 
entering into the algebra as the generators of a $U(1)$ (super) gauge 
symmery of the underlying theory on superspace. These superalgebras are gauge 
invariant, though not closed in general. We then show that these type of superalgebras are
naturally realized in a supersymmetric field theory possessing a super $U(1)$ gauge symmetry.
As the non-linearly realized symmetries of this theory, the generalized noncommutative
(super)translations and super gauge transformations are found to form a closed algebra.

\end{abstract}

\end{titlepage}

\section{\protect\smallskip \protect\smallskip Introduction}

Deformation of the supersymmetry algebras in noncommutative spaces, or more
generally in string theory in a non-trivial background, has drawn some
attentions in the context of noncommutative field theories \cite{0}-\cite{7}
(for a review on NCFT see \cite{9}, \cite{10}). In particular, in this
direction, the deformation of the $\mathcal{N}=1,D=2$ super Euclidean ($E^{2}
$) algebra, as the simplest example of these algebras, has been studied in a
recent paper \cite{6}. The motivation of that work for looking for such a
deformation was that, unlike in the ordinary case, translations of a
noncommutative space form a \textit{noncommutative} algebra \cite{8}. So one
naturally expects a corresponding deformation of the SUSY algebra when
lifting this fact to the level of a noncommutative superspace \cite{2,3}.

\smallskip

One of the main consequences of the above work was that a consistent
deformation of a SUSY algebra associated to the noncommutativity of space
naturally requires introduction of new bosonic and fermionic operators among
the usual (super)translation (and rotation) generators. It was pointed out
that these generators change \textit{only} the states in the \textit{%
fundamental} representation of the noncommutative super $E^{2}$ group \cite
{6}. As such, the deformation proposed in that (and also in this) paper has
not a direct reflection on supersymmetric field theories on noncommutative
spaces \cite{0}, because in those theories field operators naturally
transform as in the \textit{adjoint} representation \cite{13}. Indeed, such
a deformation is manifestly realized in a \textit{commutative} theory whose
field content belongs to the fundamental representation of the corresponding
supergroup.

\smallskip

It was found that the new generators in the superalgebra of the
noncommutative $E^{2}$ (briefly denoted as $E_{\vartheta }^{2}$) form a
complete ``basis'' of expansion for functions of the grassmann coordinates $%
(\theta ^{+},\theta ^{-})$. This proposes the idea that these new operators
must be interpreted as the generators of a ``local'' transformation acting
on the ``state superfield'' $S(x,\theta )$ as follows 
\[
S\rightarrow f(\theta )S, 
\]
with $f$ being any complex valued function of the grassmann coordinates.
This idea has an immediate generalization if we take $f$ to be a function of
both $x$ and $\theta $. So we are led to introduce a $U(1)$ \textit{super}
gauge transformation acting on a complex scalar superfield carrying this $%
U(1)$'s charge. A theory with this $U(1)$ gauge symmetry necessarily
involves a gauge superfield $\mathcal{A}_{\alpha }(x,\theta )$ which plays
the role of a gauge connection on a flat superspace. This connection in turn
modifies the algebra of the supercharges in a way that they close only with
the generators of this $U(1)$ super gauge transformations. In this paper we
will use this idea for deriving a generalized version of the previous NC
super $E^{2}$ algebra which provides also a unified framework for many other
generalizations of the ordinary superalgebra such as the one with a central
charge.

\smallskip

The paper is organized as follows: in section 2 we will present the
definition of a super gauge symmetry and introduce the associated gauge
superfield and show its application in generalization of the ordinary
superalgebra together with two of its important special cases. Then in
section 3 we study particular examples of this algebra by specializing the
general form of the gauge superfield and investigate about the closure of
the resulting algebras. Section 4 is devoted to the extension of the results
of section 3 to the case of a general gauge superfield configuration. In
section 5 we consider the issue of realization of this generalized type of
superalgebras in 2D field theories possessing not only an ordinary
supersymmetry but also a super gauge symmetry. In section 6 a purely bosonic
analogue of the above models, involving only the translation sector of the
complete supertranslation algebra, is constructed. The ideas of this purely
bosonic case are further applied to the supersymmetric case in section 7 to
show the closure of the general algebra, when interpreted as the algebra of
the (non-linear) superfield transformations. We conclude the paper by a
summary with some remarks in section 8.

\smallskip

\smallskip

\section{\protect\smallskip Generalization of SUSY Using a Gauge Superfield}

It is a well known fact that the noncommutativity of space in NCFT has a
description in terms of a (suitably chosen) gauge field background on a
commutative space \cite{8,9}. One is tempted to further generalize this idea
by introducing the concept of a gauge superfield background and its
associated (super) gauge transformation for theories defined on a \textit{%
commutative} superspace. For theories in $D=2,\,\mathcal{N}=1$ superspace,
we define a gauge superfield by a grassmann \textit{odd} spinor superfield $%
\mathcal{A}_{\alpha }(x,\theta )$ ($\alpha =\pm $) with a proper gauge
transformation as dictated by the algebra as follows. The two components of
this spinor are related by conjugation, 
\begin{equation}
(\mathcal{A}_{+}^{{}})^{\dagger }=-\mathcal{A}_{-}\text{ .}  \label{1}
\end{equation}
The multiplet defined by $\mathcal{A}_{\alpha }$ consists of a (complex)
scalar, a spinor and a vector field, which can be read easily from its
expansion in powers of $\theta ^{\pm }$.

Let us denote the ordinary (undeformed) SUSY generators by $\mathcal{Q}%
_{\alpha }$ (see Appendix for our notations), 
\begin{equation}
\mathcal{Q}_{\pm }=\frac{\partial }{\partial \theta ^{\pm }}+i\theta ^{\pm
}\partial _{\pm }\text{,}  \label{2}
\end{equation}

This is a linear differential operator obeying the usual Leibnitz rule, i.e. 
\begin{equation}
\mathcal{Q}_{\pm }(fg)=(\mathcal{Q}_{\pm }f)g+(-1)^{\deg (f)}f(\mathcal{Q}%
_{\pm }g),  \label{3}
\end{equation}
for any (odd or even) pair of superfields $f,g$. Then we define the deformed
generators $Q_{\alpha }$ through introduction of $\mathcal{A}_{\alpha }$ as
some \textit{superspace gauge connection }(in a particular representation)
as follows 
\begin{equation}
Q_{\alpha }=\mathcal{Q}_{\alpha }-i\mathcal{A}_{\alpha }\text{.}  \label{4}
\end{equation}
The ordinary algebra of $\mathcal{Q}_{\pm }$ together with the grassmann
property $\{\mathcal{A}_{\alpha },\mathcal{A}_{\beta }\}=0$ then give rise
to the algebra of $Q_{\pm }$ 
\begin{eqnarray}
\{Q_{+},Q_{-}\} &=&-i(\mathcal{Q}_{+}\mathcal{A}_{-}+\mathcal{Q}_{-}\mathcal{%
A}_{+}),  \nonumber \\
Q_{\pm }^{2} &=&i(\partial _{\pm }-\mathcal{Q}_{\pm }\mathcal{A}_{\pm })%
\text{.}  \label{5}
\end{eqnarray}
where on the RHS, $\mathcal{Q}_{\pm }$ acts on $\mathcal{A}_{\pm }$ as a
differential operator on a superfield. \smallskip The expression on the RHS
of the anticommutator of $Q_{+},Q_{-}$ is evidently invariant under the
following (super)gauge transformations 
\begin{equation}
\mathcal{A}_{\alpha }\rightarrow \mathcal{A}_{\alpha }+\mathcal{Q}_{\alpha
}\Lambda .  \label{6}
\end{equation}
Here $\Lambda (x,\theta )$ is an arbitrary (parity even) real scalar
superfield. This is a local gauge transformation in the sense that its
parameter $\Lambda $ is a function of the superspace coordinates. The gauge
invariant quantity on the RHS of the first eq.(\ref{5}) is called the 
\textit{superfield strength }of $\mathcal{A}_{\alpha }$ which in this paper
is denoted by $T$; 
\begin{equation}
T\equiv -i(\mathcal{Q}_{+}\mathcal{A}_{-}+\mathcal{Q}_{-}\mathcal{A}_{+}).
\label{7}
\end{equation}
Also the second eq.(\ref{5}), compared to its counterpart in the ordinary
superalgebra (first eq.(\ref{138})), hints on defining the generalized
momenta as follows 
\begin{equation}
P_{\pm }\equiv -i(\partial _{\pm }-\mathcal{Q}_{\pm }\mathcal{A}_{\pm })%
\text{.}  \label{8}
\end{equation}

\smallskip Note that the above representation of $P$, $Q$ clearly depends on
the gauge in which $\mathcal{A}_{\pm }$ is written. However, the algebra
itself comes out to be gauge invariant! (see below).

\smallskip

\subsection{Some Special Cases}

Before studying further the general case, it is worthwhile to have a look at
two of its important special cases.

\smallskip

\smallskip \textbf{Case 1: }SUSY\ with a central charge (CE-SUSY)

Let us take an $x$-independent configuration as 
\begin{equation}
\mathcal{A}_{\pm }(x,\theta )=c_{\pm }\theta ^{\mp },  \label{9}
\end{equation}
with $c_{\pm }$ being complex numbers ($c_{+}^{*}=-c_{-}$). Then 
\begin{eqnarray}
\mathcal{Q}_{+}\mathcal{A}_{+} &=&0,\qquad \mathcal{Q}_{+}\mathcal{A}%
_{-}=c_{-},  \nonumber \\
\mathcal{Q}_{-}\mathcal{A}_{+} &=&c_{+},\text{\qquad }\mathcal{Q}_{-}%
\mathcal{A}_{-}=0,  \label{10}
\end{eqnarray}
using which in the general expressions we find 
\begin{eqnarray}
\{Q_{+},Q_{-}\} &=&-i(c_{+}+c_{-}),  \nonumber \\
Q_{\pm }^{2} &=&i\partial _{\pm }.  \label{11}
\end{eqnarray}
This is the ordinary superalgebra modified by a central charge $a\equiv
-i(c_{+}+c_{-})$ (see \cite{11})

\smallskip

\smallskip

\textbf{Case 2: }SUSY\ in a noncommutative space (NC-SUSY)

Now consider the case with the background defined by 
\begin{equation}
\mathcal{A}_{\pm }(x,\theta )=\mp \omega x^{\mp }\theta ^{\pm }.  \label{12}
\end{equation}
For this configuration we have 
\begin{eqnarray}
\mathcal{Q}_{+}\mathcal{A}_{+} &=&-\omega x^{-},\qquad \mathcal{Q}_{+}%
\mathcal{A}_{-}=i\omega \theta ^{+}\theta ^{-},  \nonumber \\
\mathcal{Q}_{-}\mathcal{A}_{+} &=&i\omega \theta ^{+}\theta ^{-},\text{%
\qquad }\mathcal{Q}_{-}\mathcal{A}_{-}=\omega x^{+},  \label{13}
\end{eqnarray}
from which we find 
\begin{eqnarray}
\{Q_{+},Q_{-}\} &=&2\omega \theta ^{+}\theta ^{-},  \nonumber \\
Q_{\pm }^{2} &=&i(\partial _{\pm }\pm \omega x^{\mp }).  \label{14}
\end{eqnarray}
This is the supertranslation part of the algebra referred to as a
``noncommutative superalgebra'' in the earlier work \cite{6}.

\smallskip

\subsection{The General Case}

From the above examples we conclude that several interesting cases are
capable of a uniform description provided by introducing a gauge connection
on superspace as in the eq.(\ref{4}). Let us now return to the case of a
general gauge superfield and study its associated supertranslation algebra,
that is the algebra of the bosonic and fermionic generators $P_{\pm },Q_{\pm
}$. For this aim the only things needed are the ordinary algebra of $%
\mathcal{P}_{\pm }$, $\mathcal{Q}_{\pm }$(eqs.(\ref{138})) together with the
following simple rules 
\begin{eqnarray}
\lbrack P_{\pm },\mathcal{F}] &=&-i\partial _{\pm }\mathcal{F}  \nonumber \\
\lbrack Q_{\pm },\mathcal{F]}_{\deg \mathcal{F}} &=&_{{}}\mathcal{Q}_{\pm }%
\mathcal{F}  \label{15}
\end{eqnarray}
Here $\mathcal{F=}$ $\mathcal{F}$ $(x,\theta )$ denotes an arbitrary
superfield with even or odd grassmann parity. In the second equation, the
notation $[.,.\mathcal{]}_{\deg \mathcal{F}}$ is a commutator or
anticommutator depending on that $\mathcal{F}$ is even or odd. Using these
rules, for instance, for the commutator of $P_{+},Q_{+}$ we find 
\begin{eqnarray}
\lbrack P_{+},Q_{+}] &=&[-i(\partial _{+}-\mathcal{Q}_{+}\mathcal{A}_{+}%
\mathcal{)},\,\mathcal{Q}_{+}-i\mathcal{A}_{+}]  \nonumber \\
&=&-(\partial _{+}\mathcal{A}_{+}+i\mathcal{Q}_{+}^{2}\mathcal{A}_{+}) 
\nonumber \\
&=&-(\partial _{+}\mathcal{A}_{+}+i^{2}\partial _{+}\mathcal{A}_{+})=0.
\label{16}
\end{eqnarray}

\smallskip By similar calculations we find that the non-trivial
(anti)commutators between $P_{\pm },Q_{\pm }$ are written as 
\begin{eqnarray}
\{Q_{+},Q_{-}\} &=&-i(\mathcal{Q}_{+}\mathcal{A}_{-}+\mathcal{Q}_{-}\mathcal{%
A}_{+}),  \nonumber \\
\lbrack P_{+},P_{-}] &=&\partial _{+}\mathcal{Q}_{-}\mathcal{A}_{-}-\partial
_{-}\mathcal{Q}_{+}\mathcal{A}_{+},  \nonumber \\
\lbrack P_{+},Q_{-}] &=&-(\partial _{+}\mathcal{A}_{-}-i\mathcal{Q}_{+}%
\mathcal{Q}_{-}\mathcal{A}_{+}),  \nonumber \\
\lbrack P_{-},Q_{+}] &=&-(\partial _{-}\mathcal{A}_{+}-i\mathcal{Q}_{-}%
\mathcal{Q}_{+}\mathcal{A}_{-}).  \label{17}
\end{eqnarray}
which must be supplemented by the usual equations $Q_{\pm }^{2}=-P_{\pm }$.
The interesting point regarding the above equations is that despite $P,\,Q$
's themselves, which have gauge dependent representations, their algebra is
invariant under the gauge transformations. This is because all the
expressions on the RHS of these equations are gauge invariant combinations
of $\mathcal{A}_{\pm }$. The combination in the first line equation is
manifestly invariant while the invariance of the other three combinations is
easily proved. For example, the second combination under a gauge
transformation $\Lambda $ changes as 
\begin{eqnarray}
\delta _{\Lambda }(\partial _{+}\mathcal{Q}_{-}\mathcal{A}_{-}-\partial _{-}%
\mathcal{Q}_{+}\mathcal{A}_{+}) &=&\partial _{+}\mathcal{Q}_{-}(\mathcal{Q}%
_{-}\Lambda )-\partial _{-}\mathcal{Q}_{+}(\mathcal{Q}_{+}\Lambda ) 
\nonumber \\
&=&\partial _{+}(i\partial _{-}\Lambda )-\partial _{-}(i\partial _{+}\Lambda
)=0.  \label{18}
\end{eqnarray}
A similar conclusion holds for the other three combinations of $\mathcal{A}%
_{\pm }$. As such, we see that several representations of $P,\,Q$ 's
corresponding to different gauges of $\mathcal{A}_{\pm }$ lead to the same
algebra. Such representations are called\textit{\ gauge equivalent.}

\smallskip

\smallskip

\section{More special cases}

Let us now focus on two special cases which are straightforward
generalizations of the two cases in the previous section and are easy to
investigate about such questions as the closure of the resulting algebra.

\smallskip

\smallskip \textbf{Case 3:} (NC-SUSY with local noncommutativity parameter) 
\begin{equation}
\mathcal{A}_{\pm }(x,\theta )=\theta ^{\pm }A_{\pm }(x).  \label{19}
\end{equation}

where $A_{\pm }$ (related as $A_{+}^{*}=-A_{-}$) are components of a vector
field. Simple computations like the previous ones using this gauge field
gives 
\begin{eqnarray}
\lbrack P_{+},P_{-}] &=&F(x),  \nonumber \\
\{Q_{+},Q_{-}\} &=&\theta ^{+}\theta ^{-}F(x),  \nonumber \\
\lbrack P_{+},Q_{-}] &=&-\theta ^{-}F(x),  \nonumber \\
\lbrack P_{-},Q_{+}] &=&\theta ^{+}F(x),  \label{20}
\end{eqnarray}
besides the other trivial relations including $Q_{\pm }^{2}=-P_{\pm }$. Here 
$F(x)\equiv \partial _{+}A_{-}-\partial _{-}A_{+}$ is the field strength of
the ordinary gauge field $A_{\pm }(x)$. The fact that the two functions $%
A_{\pm }(x)$ appear in the algebra only through a particular combination, $%
F(x)$, is a reflection of the gauge invariance of the algebra. The case of
an ordinary NC space is a particular case of this in which the ordinary
field strength is a constant (proportional to the inverse of the NC scale 
\cite{6}). Hence we interpret $F(x)$ as some local (inverse) NC parameter.
The four gauge invariant quantities in the general theory therefore reduce
to a single one $F(x)$ in this special case. To examine the closure of the
algebra, let us denote these gauge invariant operators by 
\begin{equation}
F\equiv F(x),\qquad O_{\pm }\equiv \theta ^{\mp }F(x),\qquad T\equiv \theta
^{+}\theta ^{-}F(x).  \label{21}
\end{equation}
Note that $F,T$ are hermitian operators while $(O_{+})^{\dagger }=O_{-}$ .
Next we form the (anti)commutators of these operators with $P,Q$ 's: 
\begin{eqnarray}
\lbrack P_{\pm },F] &=&-i\partial _{\pm }F(x),  \nonumber \\
\{Q_{\pm },F\} &=&i\theta ^{\pm }\partial _{\pm }F(x),  \nonumber \\
&&  \nonumber \\
\lbrack P_{\pm },O_{+}] &=&-i\theta ^{-}\partial _{\pm }F(x),  \nonumber \\
\{Q_{+},O_{+}\} &=&i\theta ^{+}\theta ^{-}\partial _{+}F(x),  \nonumber \\
\{Q_{-},O_{+}\} &=&F(x),  \nonumber \\
&&  \nonumber \\
\lbrack P_{\pm },O_{-}] &=&-i\theta ^{+}\partial _{\pm }F(x),  \nonumber \\
\{Q_{+},O_{-}\} &=&F(x),  \nonumber \\
\{Q_{-},O_{-}\} &=&i\theta ^{-}\theta ^{+}\partial _{-}F(x),  \nonumber \\
&&  \nonumber \\
\lbrack P_{\pm },T] &=&-i\theta ^{+}\theta ^{-}\partial _{\pm }F(x), 
\nonumber \\
\lbrack Q_{\pm },T] &=&\pm \theta ^{\mp }F(x).  \label{22}
\end{eqnarray}
For being a closed super Lie algebra, all functions on the RHS of this
algebra must be written as linear combinations of $F(x),\,\theta ^{\pm
}F(x),\,\theta ^{+}\theta ^{-}F(x)$ which are representations for the
operators $F,\,O_{\mp },T$ in this case. This is possible if and only if 
\begin{equation}
\partial _{\pm }F(x)=\kappa _{\pm }F(x),  \label{23}
\end{equation}
for some constant $\kappa _{\pm }\in \Bbb{C}$. It is solved to 
\begin{equation}
F(x)=F_{0}\text{exp}(\kappa _{+}x^{+}+\kappa _{-}x^{-}),  \label{24}
\end{equation}
where $F_{0}$ is another constant and hermiticity of $F$ implies that 
\begin{equation}
F_{0}^{*}=F_{0},\qquad \kappa _{+}^{*}=\kappa _{-}.  \label{25}
\end{equation}
We note that the only rotationally symmetric solution is the one for $\kappa
_{\pm }=0$ corresponding to the case of a constant field strength $F=F_{0}\,$%
(the NC-SUSY case).

\smallskip

The full algebra which is now closed is found by putting the above solution
into the expressions for the (anti)commutators found in the above. The
result is as follows 
\begin{eqnarray}
\lbrack P_{+},P_{-}] &=&F,\qquad \{Q_{+},Q_{-}\}=T,\qquad [P_{\pm },Q_{\mp
}]=\mp O_{\pm },  \nonumber \\
\lbrack P_{\pm },F] &=&-i\kappa _{\pm }F,\qquad [Q_{\pm },F]=i\kappa _{\pm
}O_{\mp },  \nonumber \\
\lbrack P_{\pm },O_{+}] &=&-i\kappa _{\pm }O_{+},\qquad
\{Q_{+},O_{+}\}=i\kappa _{+}T,\qquad \{Q_{-},O_{+}\}=F,  \nonumber \\
\lbrack P_{\pm },O_{-}] &=&-i\kappa _{\pm }O_{-},\qquad
\{Q_{+},O_{-}\}=F,\qquad \{Q_{-},O_{-}\}=-i\kappa _{-}T,  \nonumber \\
\lbrack P_{\pm },T] &=&-i\kappa _{\pm }T,\qquad [Q_{\pm },T]=\pm O_{\pm },
\label{26}
\end{eqnarray}
besides the usual relations $Q_{\pm }^{2}=-P_{\pm }$.

\smallskip

The reduction of this algebra to the NC-SUSY algebra in the $\kappa _{\pm
}=0 $ case is almost evident, because in this case $F$ commutes with
everything else (it belongs to the center of the algebra) and by Schure
lemma we can take $F=F_{0}$, i.e. a constant times unity. The constant $%
F_{0} $ determines the inverse of the NC parameter as $\vartheta ^{-1}\equiv
\omega =\frac{1}{2}F_{0}$ \cite{6}.

One can explicitly check that the commutator of the $SO(2)$ generator $J$
with $T,F,O_{\pm }$ is \textit{not }closed unless for $\kappa _{\pm }=0$
which, in agreement with the above statement, produces the only rotationally
symmetric (i.e. the NC-SUSY) algebra in this category.

\smallskip

\smallskip \textbf{Case 4:} (CE-SUSY with a local central charge) 
\begin{equation}
\mathcal{A}_{\pm }(x,\theta )=\theta _{{}}^{\mp }f_{\pm }(x),  \label{27}
\end{equation}
\smallskip where $f_{\pm }$ (related as $f_{+}^{*}=-f_{-}$) are complex
scalar fields. The (anti)commutators of $P,Q$ 's in this case become 
\begin{eqnarray}
\{Q_{+},Q_{-}\} &=&f(x),  \nonumber \\
\lbrack P_{+},P_{-}] &=&\theta ^{+}\theta ^{-}\partial _{+}\partial _{-}f(x),
\nonumber \\
\lbrack P_{+},Q_{-}] &=&-i\theta ^{+}\partial _{+}f(x),  \nonumber \\
\lbrack P_{-},Q_{+}] &=&-i\theta ^{-}\partial _{-}f(x),  \label{28}
\end{eqnarray}
where $f\equiv -i(f_{+}+f_{-})$ is a real function. Again the appearance of
only this particular combination of the two functions $f_{\pm }(x)$ is a
result of gauge invariance of the algebra. Let us as in the previous case
identify the four functions of $(x,\theta )$ on the RHS of this algebra as
the operators $T,F,O_{\pm }$ ; i.e. 
\begin{equation}
T\equiv f(x),\qquad F\equiv \theta ^{+}\theta ^{-}\partial _{+}\partial
_{-}f(x),\qquad O_{\pm }\equiv \pm i\theta ^{\pm }\partial _{\pm }f(x)
\label{29}
\end{equation}
Then by a simple calculation we find 
\begin{eqnarray}
\lbrack P_{\pm },F] &=&i\theta ^{-}\theta ^{+}\partial _{\pm }\partial
_{+}\partial _{-}f(x),  \nonumber \\
\lbrack Q_{\pm },F] &=&\pm \theta ^{\mp }\partial _{+}\partial _{-}f(x), 
\nonumber \\
&&  \nonumber \\
\lbrack P_{\pm },O_{+}] &=&\theta ^{+}\partial _{\pm }\partial _{+}f(x), 
\nonumber \\
\{Q_{+},O_{+}\} &=&i\partial _{+}f(x),  \nonumber \\
\{Q_{-},O_{+}\} &=&\theta ^{+}\theta ^{-}\partial _{+}\partial _{-}f(x), 
\nonumber \\
&&  \nonumber \\
\lbrack P_{\pm },O_{-}] &=&-\theta ^{-}\partial _{\pm }\partial _{-}f(x), 
\nonumber \\
\{Q_{+},O_{-}\} &=&\theta ^{+}\theta ^{-}\partial _{+}\partial _{-}f(x), 
\nonumber \\
\{Q_{-},O_{-}\} &=&-i\partial _{-}f(x),  \nonumber \\
&&  \nonumber \\
\lbrack P_{\pm },T] &=&-i\partial _{\pm }f(x),  \nonumber \\
\lbrack Q_{\pm },T] &=&i\theta ^{\pm }\partial _{\pm }f(x).  \label{30}
\end{eqnarray}
By the same logic as in the previous case, the closure condition for this
algebra requires that $f(x)$ satisfies the equations 
\begin{equation}
\partial _{\pm }f(x)=\kappa _{\pm }f(x)+c_{\pm },  \label{31}
\end{equation}
where for the reality of $f$ (i.e. hermiticity of $T$) we need the constants 
$\kappa _{\pm },c_{\pm }\,$ satisfy 
\begin{equation}
\kappa _{+}^{*}=\kappa _{-},\qquad c_{+}^{*}=c_{-}.  \label{32}
\end{equation}
The integrability condition, $\partial _{+}(\partial _{-}f)=\partial
_{-}(\partial _{+}f)$, for the above equations further requires the relation 
\begin{equation}
\frac{c_{+}}{\kappa _{+}}=\frac{c_{-}}{\kappa _{-}}\text{,}  \label{33}
\end{equation}
Then the above equations are integrated to 
\begin{equation}
f(x)=a+b\exp (\kappa _{+}x^{+}+\kappa _{-}x^{-})  \label{34}
\end{equation}
where $a,b\in \Bbb{R}$ are arbitrary constants. Using this solution in the
previous expressions, the closed algebra under consideration becomes 
\begin{eqnarray}
\lbrack P_{+},P_{-}] &=&F,\qquad \{Q_{+},Q_{-}\}=T,\qquad [P_{\pm },Q_{\mp
}]=\mp O_{\pm },  \nonumber \\
\lbrack P_{\pm },F] &=&-i\kappa _{\pm }F,\qquad [Q_{\pm },F]=i\kappa _{\pm
}O_{\mp },  \nonumber \\
\lbrack P_{\pm },O_{+}] &=&-i\kappa _{\pm }O_{+},\qquad
\{Q_{+},O_{+}\}=i\kappa _{+}(T-a),\qquad \{Q_{-},O_{+}\}=F,  \nonumber \\
\lbrack P_{\pm },O_{-}] &=&-i\kappa _{\pm }O_{-},\qquad
\{Q_{+},O_{-}\}=F,\qquad \{Q_{-},O_{-}\}=-i\kappa _{-}(T-a),  \nonumber \\
\lbrack P_{\pm },T] &=&-i\kappa _{\pm }(T-a),\qquad [Q_{\pm },T]=\pm O_{\pm
}.  \label{35}
\end{eqnarray}
Obviously, for $a=0$, this is the same as the algebra obtained in the
previous case, though the generators now have a different (non-gauge
equivalent) representations! For nonzero $a$ it is just a simple central
extension of the same algebra as can be seen by putting $T^{\prime }\equiv
T-a.$

On the other hand, for $\kappa _{\pm }=0$, we recover the NC-SUSY algebra
written in new notations as 
\begin{eqnarray}
\lbrack P_{+},P_{-}] &=&F,\qquad \{Q_{+},Q_{-}\}=T,\qquad [P_{\pm },Q_{\mp
}]=\mp O_{\pm },  \nonumber \\
\{Q_{+},O_{-}\} &=&F,\qquad \{Q_{-},O_{+}\}=F,\qquad [Q_{\pm },T]=\pm O_{\pm
}.  \label{36}
\end{eqnarray}
Since in this case $F$ commutes with everything else, by Schure lemma, we
can treat it as a constant times the unity operator. To recover the previous
form of the NC-SUSY algebra in \cite{6}, it is sufficient to put $F=2\omega $
and then rescale $T,O_{\pm }$ as: $T\rightarrow 2\omega T,\,O_{\pm
}\rightarrow 2\omega O_{\pm }$.

\smallskip

There is a class of representations of the general algebra for which $T=a$
is a constant. Indeed, assuming $T$ to be proportional to the unity operator
in the above algebra, we find that for such representations we necessarily
have $F=0,\,O_{\pm }=0$. The resulting algebra is a centrally extended
version of the commutative algebra \cite{11} whose only nontrivial relation
is 
\begin{equation}
\{Q_{+},Q_{-}\}=a.  \label{37}
\end{equation}

\section{\protect\smallskip \protect\smallskip General Algebra of $%
T,F,O_{\pm }$ with $P_{\pm },Q_{\pm }$}

We are now going to generalize the above results to the case with an
arbitrary gauge superfield $\mathcal{A}_{\pm }(x,\theta )$. In accordance
with notations of the previous section we identify the operators $T,F,O_{\pm
}$ by the RHS expressions of eq.(\ref{17}) as follows 
\begin{eqnarray}
T &\equiv &-i(\mathcal{Q}_{+}\mathcal{A}_{-}+\mathcal{Q}_{-}\mathcal{A}_{+}),
\nonumber \\
O_{+} &\equiv &+(\partial _{+}\mathcal{A}_{-}-i\mathcal{Q}_{+}\mathcal{Q}_{-}%
\mathcal{A}_{+}),  \nonumber \\
O_{-} &\equiv &-(\partial _{-}\mathcal{A}_{+}-i\mathcal{Q}_{-}\mathcal{Q}_{+}%
\mathcal{A}_{-}),  \nonumber \\
F &\equiv &\partial _{+}\mathcal{Q}_{-}\mathcal{A}_{-}-\partial _{-}\mathcal{%
Q}_{+}\mathcal{A}_{+}.  \label{38}
\end{eqnarray}
\smallskip From the point of view of the $SO(2)$ rotations, $T,F$ are
scalars while $O_{\pm }$ behave as components of a spinor. As mentioned
earlier, these four quantities constitute a set of gauge invariant
operators, among which only $T$ is first order in derivatives of $\mathcal{A}%
_{\pm }$ while the others are of second order. So such, the most suitable
candidate for a strength of the superfield $\mathcal{A}_{\pm }$ is $T$, from
which an invariant action can be built.

\smallskip

\smallskip That these operators are gauge invariant can be seen more
explicitly by observing that $O_{\pm },F$ are related to $T$ by the
following simple equations 
\begin{eqnarray}
O_{\pm } &=&\pm \,\mathcal{Q}_{\pm }T,  \nonumber \\
F &=&\,\mathcal{Q}_{-}\mathcal{\,Q}_{+}T=-\mathcal{Q}_{+}\,\mathcal{Q}_{-}T.
\label{39}
\end{eqnarray}
These are nothing but a consequence of the algebra of $\partial _{\pm },\,%
\mathcal{Q}_{\pm }$ in the above definitions of these quantities. Using
these relations it is now easy to find the general form of the algebra which
is written as 
\begin{eqnarray}
\{Q_{+},Q_{-}\} &=&T,\qquad [P_{+},P_{-}]=F,\qquad [P_{\pm },Q_{\mp }]=\mp
O_{\pm },  \nonumber \\
\lbrack P_{\pm },F] &=&-i\partial _{\pm }F,\qquad [Q_{\pm },F]=i\partial
_{\pm }O_{\mp },\qquad  \nonumber \\
\lbrack P_{\pm },O_{+}] &=&-i\partial _{\pm }O_{+},\qquad
\{Q_{+},O_{+}\}=i\partial _{+}T,\qquad \{Q_{-},O_{+}\}=F,  \nonumber \\
\lbrack P_{\pm },O_{-}] &=&-i\partial _{\pm }O_{-},\qquad
\{Q_{+},O_{-}\}=F,\qquad \{Q_{-},O_{-}\}=-i\partial _{-}T,  \nonumber \\
\lbrack P_{\pm },T] &=&-i\partial _{\pm }T,\qquad [Q_{\pm },T]=\pm O_{\pm }.
\label{40}
\end{eqnarray}
This is just similar to the algebra in case 3 if one replaces $\kappa _{\pm
} $ by $\partial _{\pm }$. The requirement of the closure of this algebra
means that $\partial _{\pm }T,\,\partial _{\pm }F,\,\partial _{\pm }O^{\pm }$
must be written as linear combinations of $T,\,F,\,O_{\pm \text{.}}$, but
not of course of $P_{\pm },\,Q_{\pm }.$ In the particular cases of the
previous section, this leads to the same exponential configurations as we
found there.

\smallskip \newpage

\section{\protect\smallskip Realization of the Generalized SUSY in Field
Theories with Super $U(1)$ Gauge Symmetry}

\subsection{Preliminaries}

In this section we will consider the problem of constructing field theories
realizing the generalized SUSY algebra as an algebra underlying their
symmetries. These models turn out to be realizable by a class of
supersymmetric field theories which, in addition to having a \textit{global}
(ordinary) SUSY, they also possess a \textit{local} super gauge symmetry. We
shall call such models as the ``gauge superfield theories'' or briefly as
GSFT. Let us consider for convenience the simplest example of such models
involving a gauge multiplet defined by $\mathcal{A}_{\alpha }(x,\theta )$ as
well as a scalar multiplet defined by $S(x,\theta )$. These superfields obey
the following infinitesimal (super) gauge transformations 
\begin{eqnarray}
\delta _{\Lambda }\mathcal{A}_{\alpha } &=&\mathcal{Q}_{\alpha }\Lambda , 
\nonumber \\
\delta _{\Lambda }S &=&i\Lambda S,  \label{41}
\end{eqnarray}
with $\Lambda (x,\theta )$ being a real valued scalar superfield. These
superfields also transform under the ordinary SUSY transformations. The SUSY
transformation of $S$ is \footnote{%
In this section we denote several $\delta $-variations by a subscript on $%
\delta $ representing the appropriate generators. In the next section we
will change this notation by replacing these subscripts with symbols
denoting the corresponding transformation parameters.} 
\begin{equation}
\delta _{\mathcal{Q}}S=(\overline{\epsilon }\mathcal{Q})S.  \label{42}
\end{equation}
In order to obtain a SUSY invariant theory, the superfield $\mathcal{A}%
_{\alpha }$ must also transform under SUSY but its transformation should be
slightly different from that of $S$, because it does not directly enter into
the action but after a ``dualization'' procedure (see below).

\smallskip As is well known, for a superspace formulation of the ordinary
SUSY field theories \cite{12}, we need to introduce the concept of
supercovariant derivatives as follows 
\begin{equation}
\mathcal{D}_{\pm }\equiv \frac{\partial }{\partial \theta ^{\pm }}-i\theta
^{\pm }\partial _{\pm }.  \label{43}
\end{equation}
These supercovariant derivatives have the important property that they
anticommute with the SUSY generators 
\begin{equation}
\{\mathcal{D}_{\alpha },\mathcal{Q}_{\beta }\}=0,  \label{44}
\end{equation}
for all values of the spinor indices $\alpha ,\beta $. Using these
derivatives, for example, the kinetic Lagrangian of the complex scalar
superfield $S$ is written as 
\begin{equation}
\frac{1}{2}\mathcal{D}_{\alpha }S\mathcal{D}^{\alpha }\overline{S},
\label{45}
\end{equation}
where the spinorial index $\alpha $ is raised with $\varepsilon ^{\alpha
\beta }$ and lowered with $\varepsilon _{\alpha \beta }$. This term is
obviously invariant under the \textit{global }$U(1)$ transformations defined
by $S\rightarrow $e$^{i\Lambda }S$ but not under its local version. As in
the ordinary gauge theory, however, we can make it local by introducing a
gauge connection which in this case must be a spinor superfield $\mathcal{B}%
_{\alpha }$ transforming under the local $U(1)$ as

\begin{equation}
\mathcal{B}_{\alpha }\rightarrow \mathcal{B}_{\alpha }+\mathcal{D}_{\alpha
}\Lambda .  \label{46}
\end{equation}
This differs from the transformation property of $\mathcal{A}_{\alpha }$ in
the replacement of $\mathcal{Q}_{\alpha }$ with $\mathcal{D}_{\alpha }$.
Gauge invariant Lagrangians can then be obtained by replacing the ordinary
(super)covariant derivatives $\mathcal{D}_{\alpha }$ with the ``(super)
gauge covariant'' derivatives $D_{\alpha }$ whose effect on a superfield
with a $U(1)\,$ charge $e$ is defined as follows 
\begin{equation}
D_{\alpha }\equiv \mathcal{D}_{\alpha }-ie\mathcal{B}_{\alpha }.  \label{47}
\end{equation}
In this way the (super)gauge invariant supersymmetric Lagrangian of $S$
becomes 
\begin{equation}
\mathcal{L}_{S}=\frac{1}{2}D_{\alpha }SD^{\alpha }\overline{S}.  \label{48}
\end{equation}
Here $S$ and $\overline{S}$ are assumed to have the $U(1)$ charges $+1$ and $%
-1$, respectively. So 
\begin{equation}
D_{\alpha }S\equiv (\mathcal{D}_{\alpha }-i\mathcal{B}_{\alpha })S,\qquad
\qquad D_{\alpha }\overline{S}\equiv (\mathcal{D}_{\alpha }+i\mathcal{B}%
_{\alpha })\overline{S}.  \label{49}
\end{equation}

The above Lagrangian ($\mathcal{L}_{S}$) is manifestly supersymmetric
provided $\mathcal{B}_{\alpha }$ transforms under SUSY similarly to $S$ as
follows 
\begin{equation}
\delta _{\mathcal{Q}}\mathcal{B}_{\alpha }=(\overline{\epsilon }\mathcal{Q})%
\mathcal{B}_{\alpha }.  \label{50}
\end{equation}
Then it can be seen that $D_{\alpha }S$ transforms covariantly under the
ordinary SUSY transformations; i.e. 
\begin{eqnarray}
\delta _{\mathcal{Q}}(D_{\alpha }S) &=&-iS\delta _{\mathcal{Q}}\mathcal{B}%
_{\alpha }+D_{\alpha }(\delta _{\mathcal{Q}}S)  \nonumber \\
&=&[-i(\overline{\epsilon }\mathcal{Q})\mathcal{B}_{\alpha }+(\mathcal{D}%
_{\alpha }-i\mathcal{B}_{\alpha })(\overline{\epsilon }\mathcal{Q})]S 
\nonumber \\
&=&(\overline{\epsilon }\mathcal{Q})(\mathcal{D}_{\alpha }-i\mathcal{B}%
_{\alpha })S=(\overline{\epsilon }\mathcal{Q})D_{\alpha }S.  \label{51}
\end{eqnarray}
where we have used the eqs.(\ref{3}),(\ref{44}). This insures that any
analytic function of $S,$ $\overline{S}$ and their covariant derivatives
changes covariantly under SUSY transformations and hence can be used for
building a SUSY invariant action. However, other restrictions such as the
gauge invariance and Lorentz invariance specially constrain the form of the
physical Lagrangian. We emphasize that the gauge superfield $\mathcal{B}%
_{\alpha }$ in the Lagrangian is not actually independent of $\mathcal{A}%
_{\alpha }$ used to build the generalized SUSY algebra. In the following
subsection we will relate them by a certain dualization map.

\smallskip

\smallskip

\subsection{Realization of the Generalized SUSY in GSFT}

We are now ready to show that our generalized superalgebra derived from the
gauge dependent representations of the generators 
\begin{eqnarray}
Q_{\pm } &=&\mathcal{Q}_{\pm }-i\mathcal{A}_{\pm },  \nonumber \\
P_{\pm } &=&-i(\partial _{\pm }-\mathcal{Q}_{\pm }\mathcal{A}_{\pm }),
\label{52}
\end{eqnarray}
is realized in a GSFT. For this purpose, we first note that the
transformations generated by $Q_{\alpha }$ on a scalar superfield are
written as a combination of an ordinary SUSY and a gauge transformation as
follows 
\begin{equation}
\delta _{Q}S=(\overline{\epsilon }\mathcal{Q}-i\overline{\epsilon }\mathcal{A%
})S=\delta _{\mathcal{Q}}S+\delta _{\Lambda }S,  \label{53}
\end{equation}
where the gauge parameter $\Lambda $ is ``knitted'' into the SUSY parameter $%
\epsilon $ as 
\begin{equation}
\Lambda =\Lambda (\epsilon )\equiv -\overline{\epsilon }\mathcal{A}(x,\theta
).  \label{54}
\end{equation}
On the other hand, the invariance of the theory governing $(S,\mathcal{B}%
_{\alpha })$ requires a corresponding change of $\mathcal{B}_{\alpha }$,
which is expected (just as $\delta _{Q}S$) to be a combination of the
ordinary SUSY and a gauge transformation. This can be seen explicitly by
noting that $D_{\alpha }S$ must transform covariantly under the $Q$%
-transformations; i.e. 
\begin{equation}
\delta _{Q}(D_{\alpha }S)=\delta _{Q}[(\mathcal{D}_{\alpha }-i\mathcal{B}%
_{\alpha })S]=-iS\delta _{Q}\mathcal{B}_{\alpha }+D_{\alpha }(\overline{%
\epsilon }QS)=\overline{\epsilon }Q(D_{\alpha }S).  \label{55}
\end{equation}
Using the anticommutator of $D,Q$'s 
\begin{equation}
\{D_{\alpha },Q_{\beta }\}=-i\mathcal{D}_{\alpha }\mathcal{A}_{\beta }-i%
\mathcal{Q}_{\beta }\mathcal{B}_{\alpha },  \label{56}
\end{equation}
the last equation becomes 
\begin{equation}
\delta _{Q}(D_{\alpha }S)=\overline{\epsilon }Q(D_{\alpha }S)+iS[\overline{%
\epsilon }\mathcal{QB}_{\alpha }-\mathcal{D}_{\alpha }(\overline{\epsilon }%
\mathcal{A})-\delta _{Q}\mathcal{B}_{\alpha }]=\overline{\epsilon }%
Q(D_{\alpha }S),  \label{57}
\end{equation}
from which it follows 
\begin{equation}
\delta _{Q}\mathcal{B}_{\alpha }=\overline{\epsilon }\mathcal{QB}_{\alpha }-%
\mathcal{D}_{\alpha }(\overline{\epsilon }\mathcal{A})=\delta _{\mathcal{Q}}%
\mathcal{B}_{\alpha }+\delta _{\Lambda }\mathcal{B}_{\alpha }.  \label{58}
\end{equation}
This is the expected result which again needs relating $\Lambda $ to $%
\epsilon $ (and $\mathcal{A}$) as in the above.

\smallskip

\smallskip

Up to now, we have not assumed any relation between the two gauge
superfields $\mathcal{A}_{\alpha }$ and $\mathcal{B}_{\alpha }$, i.e. the
one appearing in the representation of the algebra and the other in gauging
of the field theory. They must be related, however, if we suppose that they
should transform under the same gauge transformation with a parameter $%
\Lambda $. Therefore, we must look for a map of the form $\mathcal{B}=%
\mathcal{B(A)}$ with the property that transforming $\mathcal{A}$ like 
\begin{equation}
\mathcal{A}_{\alpha }\mathcal{\rightarrow A}_{\alpha }\mathcal{+Q}_{\alpha
}\Lambda ,  \label{59}
\end{equation}
corresponds to a transformation of $\mathcal{B}$ as 
\begin{equation}
\mathcal{B}_{\alpha }\mathcal{\rightarrow B}_{\alpha }\mathcal{+D}_{\alpha
}\Lambda .  \label{60}
\end{equation}
\smallskip This is indeed an easy task, if we keep in mind that every
superfield $\Lambda (x,\theta )$ has a unique decomposition of the form 
\begin{equation}
\Lambda =\Lambda ^{(1)}+\Lambda ^{(2)}+\Lambda ^{(3)}+\Lambda ^{(4)},
\label{61}
\end{equation}
where $\Lambda ^{(1)},\Lambda ^{(2)},\Lambda ^{(3)},\Lambda ^{(4)}$ stand
for the terms proportional to $1,\theta ^{+},\theta ^{-},\theta ^{+}\theta
^{-}$ in the expansion of $\Lambda $, respectively. The advantage of
decomposing $\Lambda $ in this way is that the effect of each of the
operators $\mathcal{Q}_{\pm }$ or $\mathcal{D}_{\pm }$ on a term in each
class $\Lambda ^{(i)}$ is in another definite class like $\Lambda ^{(j)}$.
Therefore, we can easily relate the operations of $\mathcal{Q}$ and $%
\mathcal{D}$ on terms in each class and, as such, relate several terms of $%
\mathcal{A}_{\pm }$ to that of $\mathcal{B}_{\pm }$. The result of this
analysis is summarized in the following: 
\begin{eqnarray}
\mathcal{Q}_{+}\Lambda ^{(1)} &=&-\mathcal{D}_{+}\Lambda ^{(1)},\qquad 
\mathcal{Q}_{-}\Lambda ^{(1)}=-\mathcal{D}_{-}\Lambda ^{(1)},  \nonumber \\
\mathcal{Q}_{+}\Lambda ^{(2)} &=&\mathcal{D}_{+}\Lambda ^{(2)},\qquad 
\mathcal{Q}_{-}\Lambda ^{(2)}=-\mathcal{D}_{-}\Lambda ^{(2)},  \nonumber \\
\mathcal{Q}_{+}\Lambda ^{(3)} &=&-\mathcal{D}_{+}\Lambda ^{(3)},\qquad 
\mathcal{Q}_{-}\Lambda ^{(3)}=\mathcal{D}_{-}\Lambda ^{(3)},  \nonumber \\
\mathcal{Q}_{+}\Lambda ^{(4)} &=&\mathcal{D}_{+}\Lambda ^{(4)},\qquad 
\mathcal{Q}_{-}\Lambda ^{(4)}=\mathcal{D}_{-}\Lambda ^{(4)}.  \label{62}
\end{eqnarray}
Applying a similar decomposition to $\mathcal{A}_{\alpha },$%
\begin{equation}
\mathcal{A}_{\alpha }=\mathcal{A}_{\alpha }^{(1)}+\mathcal{A}_{\alpha
}^{(2)}+\mathcal{A}_{\alpha }^{(3)}+\mathcal{A}_{\alpha }^{(4)},  \label{63}
\end{equation}
and identifying the same class terms on both sides of $\delta _{\Lambda }%
\mathcal{A}_{\alpha }=\mathcal{Q}_{\alpha }\Lambda $, and using the above
set of equations, we finally obtain 
\begin{eqnarray}
\delta _{\Lambda }(\mathcal{A}_{+}^{(1)}-\mathcal{A}_{+}^{(2)}+\mathcal{A}%
_{+}^{(3)}-\mathcal{A}_{+}^{(4)}) &=&\mathcal{D}_{+}(\Lambda ^{(1)}+\Lambda
^{(2)}+\Lambda ^{(3)}+\Lambda ^{(4)}),  \nonumber \\
\delta _{\Lambda }(\mathcal{A}_{-}^{(1)}+\mathcal{A}_{-}^{(2)}-\mathcal{A}%
_{-}^{(3)}-\mathcal{A}_{-}^{(4)}) &=&\mathcal{D}_{-}(\Lambda ^{(1)}+\Lambda
^{(2)}+\Lambda ^{(3)}+\Lambda ^{(4)}).  \label{64}
\end{eqnarray}
This gives the identification of $\mathcal{B}_{\pm }$ in terms of the
components of $\mathcal{A}_{\pm }\,$as follows 
\begin{eqnarray}
\mathcal{B}_{+} &=&\mathcal{A}_{+}^{(1)}-\mathcal{A}_{+}^{(2)}+\mathcal{A}%
_{+}^{(3)}-\mathcal{A}_{+}^{(4)},  \nonumber \\
\mathcal{B}_{-} &=&\mathcal{A}_{-}^{(1)}+\mathcal{A}_{-}^{(2)}-\mathcal{A}%
_{-}^{(3)}-\mathcal{A}_{-}^{(4)},  \label{65}
\end{eqnarray}
upon which the desired transformation property ($\delta _{\Lambda }\mathcal{B%
}_{\alpha }=\mathcal{D}_{\alpha }\Lambda $) results. We notice that the
above map preserves the conjugation properties of the gauge superfields;
i.e. 
\begin{equation}
(\mathcal{A}_{\pm })^{\dagger }=-\mathcal{A}_{\mp }\qquad \Leftrightarrow
\qquad (\mathcal{B}_{\pm })^{\dagger }=-\mathcal{B}_{\mp }.  \label{66}
\end{equation}
It is seen that the map $\mathcal{B}=\mathcal{B(A)\,}$ is linear and hence
can be written as 
\begin{equation}
\mathcal{B}_{+}=C\mathcal{A}_{+},\qquad \mathcal{B}_{-}=\overline{C}\mathcal{%
A}_{-}.  \label{67}
\end{equation}
The operators $C,\overline{C}$ are defined by their action on an arbitrary
superfield $\Lambda (x,\theta )$ as follows 
\begin{eqnarray}
C\Lambda &\equiv &\Lambda ^{(1)}-\Lambda ^{(2)}+\Lambda ^{(3)}-\Lambda
^{(4)},  \nonumber \\
\overline{C}\Lambda &\equiv &\Lambda ^{(1)}+\Lambda ^{(2)}-\Lambda
^{(3)}-\Lambda ^{(4)}.  \label{68}
\end{eqnarray}
It is easy then to check that 
\begin{equation}
^{{}}C^{2}=\overline{C}^{2}=1,\qquad [C,\overline{C}]=0.  \label{69}
\end{equation}
Hence the inverses of $C,\overline{C}$ are given by themselves. Also the
combined effect of $C,\overline{C}$ on $\Lambda $ is given by 
\begin{equation}
C\overline{C}\Lambda =\Lambda ^{(1)}-\Lambda ^{(2)}-\Lambda ^{(3)}+\Lambda
^{(4)}.  \label{70}
\end{equation}
We can use $C,\overline{C}$ to relate $\mathcal{Q}_{\pm }$ to $\mathcal{D}%
_{\pm }$%
\begin{equation}
\mathcal{D}_{+}=C\mathcal{Q}_{+},\qquad \mathcal{D}_{-}=\overline{C}\mathcal{%
Q}_{-}.  \label{71}
\end{equation}
These relations are easily proved by applying both sides of them on an
arbitrary superfield $\Lambda (x,\theta )$. These together with the eq.(\ref
{67}) propose a ``duality'' of the form 
\begin{eqnarray}
\mathcal{A}_{\alpha } &\leftrightarrow &\mathcal{B}_{\alpha },  \nonumber \\
\mathcal{Q}_{\alpha } &\leftrightarrow &\mathcal{D}_{\alpha }.  \label{72}
\end{eqnarray}
In a similar way, we can find the following rules for the commutation of $C,%
\overline{C}$ with $\mathcal{Q}_{\pm }$%
\begin{eqnarray}
C\mathcal{Q}_{+} &=&-\mathcal{Q}_{+}C,\qquad \overline{C}\mathcal{Q}_{+}=%
\mathcal{Q}_{+}\overline{C},  \nonumber \\
C\mathcal{Q}_{-} &=&\mathcal{Q}_{-}C,\qquad \overline{C}\mathcal{Q}_{-}=-%
\mathcal{Q}_{-}\overline{C}.  \label{73}
\end{eqnarray}
By replacing $\mathcal{Q}_{\pm }$ in terms of $\mathcal{D}_{\pm }$ in these
relations, we find their dual relations 
\begin{eqnarray}
C\mathcal{D}_{+} &=&-\mathcal{D}_{+}C,\qquad \overline{C}\mathcal{D}_{+}=%
\mathcal{D}_{+}\overline{C},  \nonumber \\
C\mathcal{D}_{-} &=&\mathcal{D}_{-}C,\qquad \overline{C}\mathcal{D}_{-}=-%
\mathcal{D}_{-}\overline{C}.  \label{74}
\end{eqnarray}
As an application of these formulae, let us derive the $Q$-transformation of 
$\mathcal{B}_{\alpha }$ using that of $\mathcal{A}_{\alpha }$. We have seen
that 
\begin{equation}
\delta _{Q}\mathcal{B}_{\alpha }=\overline{\epsilon }\mathcal{QB}_{\alpha }-%
\mathcal{D}_{\alpha }(\overline{\epsilon }\mathcal{A}).  \label{75}
\end{equation}
We can express this totally in terms of $\mathcal{A}$ using the above
mentioned map between $\mathcal{A}$ and $\mathcal{B}$. In terms of the
components, this becomes 
\begin{eqnarray}
\delta _{Q}(C\mathcal{A}_{+}) &=&(\epsilon ^{+}\mathcal{Q}_{+}+\epsilon ^{-}%
\mathcal{Q}_{-})C\mathcal{A}_{+}-\mathcal{D}_{+}(\overline{\epsilon }%
\mathcal{A})  \nonumber \\
\delta _{Q}(\overline{C}\mathcal{A}_{-}) &=&(\epsilon ^{+}\mathcal{Q}%
_{+}+\epsilon ^{-}\mathcal{Q}_{-})\overline{C}\mathcal{A}_{-}-\mathcal{D}%
_{-}(\overline{\epsilon }\mathcal{A})  \label{76}
\end{eqnarray}
By multiplication of the first (second) equation by $C$ ($\overline{C}$) and
replacing $\mathcal{D}_{+}$ ($\mathcal{D}_{-}$) in terms of $\mathcal{Q}_{+}$
($\mathcal{Q}_{-}$) and using all the above algebra, we finally obtain 
\begin{equation}
\delta _{Q}\mathcal{A}_{\pm }=\epsilon ^{\mp }(\mathcal{Q}_{+}\mathcal{A}%
_{-}+\mathcal{Q}_{-}\mathcal{A}_{+})=i\epsilon ^{\mp }T.  \label{77}
\end{equation}
After all, this shows that the $Q$-variation of $\mathcal{A}_{\pm }$ is a
gauge invariant quantity! This is a general property of the $Q$%
-transformations: indeed one can check that the $Q$-transformations of $(S,%
\mathcal{A}_{\alpha })$ 
\begin{eqnarray}
\delta _{Q}S &=&\epsilon ^{\alpha }(\mathcal{Q}_{\alpha }-i\mathcal{A}%
_{\alpha })S,  \nonumber \\
\delta _{Q}\mathcal{A}_{\pm } &=&i\epsilon ^{\mp }T,  \label{78}
\end{eqnarray}
commute with their gauge transformations $\Lambda $; i.e. in general 
\begin{equation}
\lbrack \delta _{\Lambda },\delta _{Q}]=0.  \label{79}
\end{equation}
This is equivalent to saying that $\delta _{Q}$ of every superfield in the
theory is a gauge covariant (or invariant) quantity.

\smallskip

\smallskip

\section{A Purely Bosonic Analogue}

The above constructions have a purely bosonic counterpart in the ordinary
(abelian) gauge theory. Let us for simplicity work with the case of a single
charged (complex) scalar field $\phi (x)$ coupled to a gauge field $A_{\mu
}(x)$ living in an arbitrary dimensional (Minkowski) spacetime. This theory
is defined by the Lagrangian 
\begin{equation}
\mathcal{L}=-\frac{1}{4}F_{\mu \nu }F^{\mu \nu }+\frac{1}{2}\nabla _{\mu
}\phi \nabla ^{\mu }\phi ^{*},  \label{80}
\end{equation}
which is a Poincare and gauge invariant quantity. The gauge covariant
derivative $\nabla _{\mu }$ on $\phi $, $\phi ^{*}$ (with charges $\pm 1$)
is defined as usual as 
\begin{equation}
\nabla _{\mu }\phi \equiv (\partial _{\mu }-iA_{\mu })\phi ,\qquad \nabla
_{\mu }\phi ^{*}\equiv (\partial _{\mu }+iA_{\mu })\phi ^{*}.  \label{81}
\end{equation}
The above Lagrangian is invariant under the translations 
\begin{equation}
\delta _{a}\phi =a^{\nu }\partial _{\nu }\phi ,\qquad \delta _{a}A_{\mu
}=a^{\nu }\partial _{\nu }A_{\mu },  \label{82}
\end{equation}
as well as under the gauge transformations 
\begin{equation}
\delta _{\Lambda }\phi =i\Lambda \phi ,\qquad \delta _{\Lambda }A_{\mu
}=\partial _{\mu }\Lambda .  \label{83}
\end{equation}
Obviously, a general gauge transformation does not commute with translations 
\begin{equation}
\lbrack \delta _{a},\delta _{\Lambda }]\neq 0.  \label{84}
\end{equation}
However, we can generalize translations to a new (non-linearly realized)
symmetry of the theory commuting with gauge transformations. This is
obtained by ``knitting'' the (local) gauge transformations and the (global)
translations through relating their parameters in a Lorentz invariant way as 
\begin{equation}
\Lambda (a)=-a^{\mu }A_{\mu }(x).  \label{85}
\end{equation}
Let us show the combined effect of these two transformations by 
\begin{equation}
\widehat{\delta }_{a}\equiv \delta _{a}+\delta _{\Lambda (a)}.  \label{86}
\end{equation}
Then the effect of this new (global) transformation on $(\phi ,A_{\mu })$ is
written as 
\begin{eqnarray}
\widehat{\delta }_{a}\phi &=&a^{\mu }(\partial _{\mu }-iA_{\mu })\phi
=a^{\mu }\nabla _{\mu }\phi ,  \nonumber \\
\widehat{\delta }_{a}A_{\mu } &=&a^{\nu }(\partial _{\nu }A_{\mu }-\partial
_{\mu }A_{\nu })=a^{\nu }F_{\nu \mu }.  \label{87}
\end{eqnarray}
Obviously, these are written in the form of gauge covariant quantities which
implies that the transformations defined by $\widehat{\delta }_{a}$ and $%
\delta _{\Lambda }$ commute, 
\begin{equation}
\lbrack \widehat{\delta }_{a},\delta _{\Lambda }]=0.  \label{88}
\end{equation}

We now consider the commutator of two $\widehat{\delta }_{a}$'s. For
simplicity it is better to work with $\widehat{\delta }_{\mu }$'s, instead
of $\widehat{\delta }_{a}$, which are related to the latter by 
\begin{equation}
\widehat{\delta }_{a}\equiv a^{\mu }\widehat{\delta }_{\mu }.  \label{89}
\end{equation}
For the above purpose, we first compute $\widehat{\delta }_{\mu }(\widehat{%
\delta }_{\nu }\phi )$ and $\widehat{\delta }_{\mu }(\widehat{\delta }_{\nu
}A_{\rho })$%
\begin{eqnarray}
\widehat{\delta }_{\mu }(\widehat{\delta }_{\nu }\phi ) &=&\widehat{\delta }%
_{\mu }(\nabla _{\nu }\phi )=-i(\widehat{\delta }_{\mu }A_{\nu })\phi
+\nabla _{\nu }(\widehat{\delta }_{\mu }\phi )=-iF_{\mu \nu }\phi +\nabla
_{\nu }\nabla _{\mu }\phi ,  \nonumber \\
\widehat{\delta }_{\mu }(\widehat{\delta }_{\nu }A_{\rho }) &=&\widehat{%
\delta }_{\mu }F_{\nu \rho }=\partial _{\nu }(\widehat{\delta }_{\mu
}A_{\rho })-\partial _{\rho }(\widehat{\delta }_{\mu }A_{\nu })=\partial
_{\nu }F_{\mu \rho }-\partial _{\rho }F_{\mu \nu }=\partial _{\mu }F_{\nu
\rho },  \label{90}
\end{eqnarray}
where in the last step we used the Bianchi identities on $F_{\mu \nu }$.
Antisymmetrizing with respect to $\mu ,\nu $ then gives 
\begin{eqnarray}
\lbrack \widehat{\delta }_{\mu },\widehat{\delta }_{\mu }]\phi &=&-2iF_{\mu
\nu }\phi -[\nabla _{\mu },\nabla _{\nu }]\phi =-iF_{\mu \nu }\phi , 
\nonumber \\
\lbrack \widehat{\delta }_{\mu },\widehat{\delta }_{\mu }]A_{\rho }
&=&\partial _{\mu }F_{\nu \rho }-\partial _{\nu }F_{\mu \rho }=-\partial
_{\rho }F_{\mu \nu }.  \label{91}
\end{eqnarray}
where in the first line we have used of $[\nabla _{\mu },\nabla _{\nu
}]=-iF_{\mu \nu }$ and in the last line of the Bianchi identities.. The
expressions on the RHS are obviously in the form of gauge transformations on 
$(\phi ,A_{\rho })$ with a parameter equal to $-F_{\mu \nu }$. Thus for
general transformations $\widehat{\delta }_{a},\widehat{\delta }_{b}$ with
parameters $a^{\mu },b^{\mu }$ we conclude 
\begin{equation}
\lbrack \widehat{\delta }_{a},\widehat{\delta }_{b}]=\delta _{\Lambda (a,b)},
\label{92}
\end{equation}
where the field dependent parameter of the gauge transformations $\Lambda
(a,b)$ is given by 
\begin{equation}
\Lambda (a,b)=-a^{\mu }b^{\nu }F_{\mu \nu }(x).  \label{93}
\end{equation}
We notice that this $\Lambda $ itself is a gauge invariant quantity. The
algebra of the variations $\widehat{\delta }_{a},\delta _{\Lambda }$ which
is summarized by the eqs.(\ref{88}),(\ref{92}) is obviously closed. This is
in contrast to the algebra of the generators of these variations naively
obtained by taking their commutators; i.e. 
\begin{equation}
\lbrack \nabla _{\mu },\nabla _{\nu }]=-iF_{\mu \nu ,}\qquad [\nabla _{\rho
},F_{\mu \nu }]=\partial _{\rho }F_{\mu \nu }.  \label{94}
\end{equation}
This is a closed algebra only for those configurations of $F_{\mu \nu }(x)$
satisfying differential equations of the form 
\begin{equation}
\partial _{\rho }F_{\mu \nu }=C_{\rho \mu \nu }^{\kappa \lambda }F_{\kappa
\lambda }+D_{\rho \mu \nu },  \label{95}
\end{equation}
for some constant $C,D$ parameters. In particular, constant $F_{\mu \nu }$
configurations lead to a closed Heisenberg algebra.

\smallskip

\section{The Algebra of the Superfields Transformations}

In this section, we give direct generalizations of the results of the
previous section on the closure of the algebra of the fields variations to a
supersymmetrized version of it appearing in the case of a GSFT. In this case
we find that the variations of the superfields $(S,\mathcal{A}_{\alpha })$
under the non-linear ``generalized'' versions of the translations,
supersymmetry, and gauge transformations (i.e., $\delta _{a},\delta
_{\epsilon },\delta _{\Lambda }$) form a closed algebra.

\smallskip

Let us begin this analysis by computing the effect of the commutator $%
[\delta _{\epsilon },\delta _{\epsilon ^{\prime }}]$ on $S$ for two
different spinor parameters $\epsilon ,\epsilon ^{\prime }$. Initially, we
have 
\begin{equation}
\delta _{\epsilon }S=\overline{\epsilon }QS=(\overline{\epsilon }\mathcal{Q}%
+i\Lambda (\epsilon ))S,  \label{96}
\end{equation}
where $\Lambda (\epsilon )$ is defined by the eq.(\ref{54}). The $\delta
_{\epsilon ^{\prime }}$ variation of this expression gives 
\begin{equation}
\delta _{\epsilon ^{\prime }}(\delta _{\epsilon }S)=iS\delta _{\epsilon
^{\prime }}\Lambda (\epsilon )+(\overline{\epsilon }Q)(\overline{\epsilon }%
^{\prime }Q)S.  \label{97}
\end{equation}
Now, we have 
\begin{equation}
\delta _{\epsilon ^{\prime }}\Lambda (\epsilon )=-\delta _{\epsilon ^{\prime
}}(\epsilon ^{+}\mathcal{A}_{+}+\epsilon ^{-}\mathcal{A}_{-})=-(\epsilon
^{+}\epsilon ^{\prime -}+\epsilon ^{-}\epsilon ^{\prime +})iT,  \label{98}
\end{equation}
where we used the previous result for $\delta _{\epsilon }\mathcal{A}_{\pm }$%
, eq.(\ref{77}). Note that $\delta _{\epsilon ^{\prime }}\Lambda (\epsilon
)=-\delta _{\epsilon }\Lambda (\epsilon ^{\prime }).$ As a result, we obtain 
\begin{equation}
\lbrack \delta _{\epsilon },\delta _{\epsilon ^{\prime }}]S=-[\overline{%
\epsilon }Q,\overline{\epsilon }^{\prime }Q]S-2(\epsilon ^{+}\epsilon
^{\prime -}+\epsilon ^{-}\epsilon ^{\prime +})TS.  \label{99}
\end{equation}
The first term in this equation is easily computed using the algebra of $Q$%
's, eq.(\ref{5}), as follows 
\begin{eqnarray}
\lbrack \overline{\epsilon }Q,\overline{\epsilon }^{\prime }Q] &=&-2\epsilon
^{+}\epsilon ^{\prime +}Q_{+}^{2}-2\epsilon ^{-}\epsilon ^{\prime
-}Q_{-}^{2}-(\epsilon ^{+}\epsilon ^{\prime -}+\epsilon ^{-}\epsilon
^{\prime +})\{Q_{+},Q_{-}\}  \nonumber \\
&=&2\epsilon ^{+}\epsilon ^{\prime +}P_{+}+2\epsilon ^{-}\epsilon ^{\prime
-}P_{-}-(\epsilon ^{+}\epsilon ^{\prime -}+\epsilon ^{-}\epsilon ^{\prime
+})T.  \label{100}
\end{eqnarray}
Using this in the previous equation gives finally 
\begin{equation}
\lbrack \delta _{\epsilon },\delta _{\epsilon ^{\prime }}]S=-2(\epsilon
^{+}\epsilon ^{\prime +}P_{+}+\epsilon ^{-}\epsilon ^{\prime
-}P_{-})S-(\epsilon ^{+}\epsilon ^{\prime -}+\epsilon ^{-}\epsilon ^{\prime
+})TS.  \label{101}
\end{equation}
The RHS of this equation is evidently written as a combination of a
generalized translation (generated by $P_{\pm }$) with a gauge
transformation, i.e. 
\begin{equation}
\lbrack \delta _{\epsilon },\delta _{\epsilon ^{\prime }}]S=\delta
_{a}S+\delta _{\Lambda }S\equiv i(a^{+}P_{+}+a^{-}P_{-})S+i\Lambda S,
\label{102}
\end{equation}
where the parameters $a^{\pm },\Lambda $ of these two transformations depend
on $(\epsilon ,\epsilon ^{\prime })$ as follows 
\begin{eqnarray}
a^{\pm }(\epsilon ,\epsilon ^{\prime }) &\equiv &2i\epsilon ^{\pm }\epsilon
^{\prime \pm },  \nonumber \\
\Lambda (\epsilon ,\epsilon ^{\prime }) &\equiv &i(\epsilon ^{+}\epsilon
^{\prime -}+\epsilon ^{-}\epsilon ^{\prime +})T.  \label{103}
\end{eqnarray}
We note that the parameter of gauge transformations $\Lambda (x,\theta )$ is
a gauge invariant quantity proportional to $T(x,\theta )$. We now consider
the effect of $[\delta _{\epsilon },\delta _{\epsilon ^{\prime }}]$ on $%
\mathcal{A}_{\alpha }.$ Taking the $\delta _{\epsilon ^{\prime }\text{ }}$%
-variation of $\delta _{\epsilon }\mathcal{A}_{\pm }$, we find 
\begin{eqnarray}
\delta _{\epsilon ^{\prime }}(\delta _{\epsilon }\mathcal{A}_{\pm })
&=&\delta _{\epsilon ^{\prime }}(i\epsilon ^{\mp }T)=\epsilon ^{\mp }[%
\mathcal{Q}_{+}(\delta _{\epsilon ^{\prime }}\mathcal{A}_{-})\mathcal{+Q}%
_{-}(\delta _{\epsilon ^{\prime }}\mathcal{A}_{+})]  \nonumber \\
&=&-i\epsilon ^{\mp }(\epsilon ^{\prime +}\mathcal{Q}_{+}+\epsilon ^{\prime
-}\mathcal{Q}_{-})T.  \label{104}
\end{eqnarray}
Antisymmetrizing this expression with respect to $(\epsilon ,\epsilon
^{\prime })$, we obtain 
\begin{equation}
\lbrack \delta _{\epsilon },\delta _{\epsilon ^{\prime }}]\mathcal{A}_{\pm
}=2i\epsilon ^{\mp }\epsilon ^{\prime \mp }\mathcal{Q}_{\mp }T+i(\epsilon
^{+}\epsilon ^{\prime -}+\epsilon ^{-}\epsilon ^{\prime +})\mathcal{Q}_{\pm
}T.  \label{105}
\end{equation}
The second term on the RHS is the gauge variation of $\mathcal{A}_{\pm }$
with a parameter $\Lambda (\epsilon ,\epsilon ^{\prime })$ as defined for $S$
by eq.(\ref{103}). The first term which is proportional to $a^{\mp
}(\epsilon ,\epsilon ^{\prime })$ is the \textit{definition }of $\delta _{a}%
\mathcal{A}_{\pm }$ (see also below) 
\begin{equation}
\delta _{a}\mathcal{A}_{\pm }\equiv a^{\mp }\mathcal{Q}_{\mp }T.  \label{106}
\end{equation}
As such, the last equation is in the same form as expected from the eq.(\ref
{102}); i.e. 
\begin{equation}
\lbrack \delta _{\epsilon },\delta _{\epsilon ^{\prime }}]\mathcal{A}_{\pm
}=\delta _{a}\mathcal{A}_{\pm }+\delta _{\Lambda }\mathcal{A}_{\pm }.
\label{107}
\end{equation}

\smallskip

Note that the above definition for $\delta _{a}\mathcal{A}_{\pm }$ is
consistent with what we may expect from the purely bosonic theory introduced
in the last section. Indeed , the purely bosonic subalgebra of the
generalized SUSY algebra underlying the GSFT is the one obtained from $%
P_{\pm }$ using their representations by 
\begin{equation}
P_{\pm }\equiv -i\nabla _{\pm }\equiv -i(\partial _{\pm }-iA_{\pm }),
\label{108}
\end{equation}
where $A_{\pm }$ is defined as 
\begin{equation}
A_{\pm }\equiv -i\mathcal{Q}_{\pm }\mathcal{A}_{\pm }.  \label{109}
\end{equation}
\smallskip This mimics the ordinary gauge field $A_{\mu }(x)$ in the bosonic
theory, though here $A_{\pm }(x,\theta )$ is not a field but a superfield.
Then from the bosonic theory the $\delta _{a}$-variations (there denoted as $%
\widehat{\delta }_{a}$) of $S,A_{\pm }$ become 
\begin{eqnarray}
\delta _{a}S &=&a^{+}\nabla _{+}S+a^{-}\nabla _{-}S,  \nonumber \\
\delta _{a}A_{\pm } &=&\pm ia^{\mp }F,  \label{110}
\end{eqnarray}
where $F$ is the analogue of the ordinary field strength; i.e. 
\begin{equation}
F\equiv i(\partial _{+}A_{-}-\partial _{-}A_{+})=\partial _{+}\mathcal{Q}_{-}%
\mathcal{A}_{-}-\partial _{-}\mathcal{Q}_{+}\mathcal{A}_{+}.  \label{111}
\end{equation}
Assuming now the previous definition for $\delta _{a}\mathcal{A}_{\pm }$, we
see that $\delta _{a}A_{\pm }$ takes precisely the expected form 
\begin{equation}
\delta _{a}A_{\pm }=-i\mathcal{Q}_{\pm }(\delta _{a}\mathcal{A}_{\pm })=-i%
\mathcal{Q}_{\pm }(a^{\mp }\mathcal{Q}_{\mp }T)=\pm ia^{\mp }F,  \label{112}
\end{equation}
where we have used the relation $F\equiv \mathcal{Q}_{-}\mathcal{Q}_{+}T.$

\smallskip

\smallskip

\smallskip

Similar to the purely bosonic case, the gauge invariances of the $\delta
_{a} $-variations of the superfields (as seen from the eqs.(\ref{106}),(\ref
{110})) imply that these variations commute with their gauge variations,
i.e. 
\begin{equation}
\lbrack \delta _{\Lambda },\delta _{a}]S=[\delta _{\Lambda },\delta _{a}]%
\mathcal{A}_{\pm }=0.  \label{113}
\end{equation}
A similar statement holds for $\delta _{\epsilon }$, as we have seen 
\begin{equation}
\lbrack \delta _{\Lambda },\delta _{\epsilon }]S=[\delta _{\Lambda },\delta
_{\epsilon }]\mathcal{A}_{\pm }=0.  \label{114}
\end{equation}

\smallskip

\smallskip Let us now consider the commutator of two $\delta _{a}$'s.
Firstly, for $\mathcal{A}_{\pm }$ we have 
\begin{eqnarray}
\delta _{a}(\delta _{a^{\prime }}\mathcal{A}_{\pm }) &=&\delta
_{a}(a^{\prime \mp }\mathcal{Q}_{\mp }T)  \nonumber \\
&=&-ia^{\prime \mp }\mathcal{Q}_{\mp }(\mathcal{Q}_{+}a^{+}\mathcal{Q}_{+}T+%
\mathcal{Q}_{-}a^{-}\mathcal{Q}_{-}T)  \nonumber \\
&=&a^{\prime \mp }\mathcal{Q}_{\mp }(a^{+}\partial _{+}+a^{-}\partial _{-})T,
\label{115}
\end{eqnarray}
where we used repeatedly of the eq.(\ref{106}) and the definition of $T.$ As
a result we obtain 
\begin{equation}
\lbrack \delta _{a},\delta _{a^{\prime }}]\mathcal{A}_{\pm }=\pm
(a^{+}a^{\prime -}-a^{-}a^{\prime +})\partial _{\pm }\mathcal{Q}_{\mp }T=%
\mathcal{Q}_{\pm }[i(a^{+}a^{\prime -}-a^{-}a^{\prime +})\mathcal{Q}_{-}%
\mathcal{Q}_{+}T],  \label{116}
\end{equation}
where we used some algebra of $\mathcal{Q}$'s. We see that the commutator of
two $\delta _{a}$'s is a gauge transformation with a parameter 
\begin{equation}
\Lambda (a,a^{\prime })\equiv i(a\wedge a^{\prime })\mathcal{Q}_{-}\mathcal{Q%
}_{+}T=i(a\wedge a^{\prime })F.  \label{117}
\end{equation}
where $a\wedge a^{\prime }\equiv (a^{+}a^{\prime -}-a^{-}a^{\prime +}).$
Note that the gauge parameter itself is a gauge invariant quantity
proportional to $F$. To be complete, we prove the same statement also for $S$%
. We have 
\begin{eqnarray}
\delta _{a}(\delta _{a^{\prime }}S) &=&\delta _{a}(a^{\prime +}\nabla
_{+}S+a^{\prime -}\nabla _{-}S)  \nonumber \\
&=&a^{\prime +}(\nabla _{+}\delta _{a}S-iS\delta _{a}A_{+})+a^{\prime
-}(\nabla _{-}\delta _{a}S-iS\delta _{a}A_{-}).  \label{118}
\end{eqnarray}
Applying the previous expressions for $\delta _{a}S$ and $\delta _{a}A_{\pm
} $ to this equation and using the algebra of $\nabla _{\pm }$, we find 
\begin{eqnarray}
\lbrack \delta _{a},\delta _{a^{\prime }}]S &=&-(a^{+}a^{\prime
-}-a^{-}a^{\prime +})[\nabla _{+},\nabla _{-}]S-2(a^{+}a^{\prime
-}-a^{-}a^{\prime +})FS  \nonumber \\
&=&(a\wedge a^{\prime })FS-2(a\wedge a^{\prime })FS=i\Lambda (a,a^{\prime
})S.  \label{119}
\end{eqnarray}
This is the expected gauge transformation of $S$ with the same parameter as
found for $\mathcal{A}_{\pm }$.

\smallskip

\smallskip

Finally, we consider the commutator of $\delta _{\epsilon }$ and $\delta
_{a} $ on $S,\mathcal{A}_{\pm }$. On $\mathcal{A}_{\pm }$ we have 
\begin{eqnarray}
\delta _{\epsilon }(\delta _{a}\mathcal{A}_{\pm }) &=&a^{\mp }\mathcal{Q}%
_{\mp }(\delta _{\epsilon }T)=-a^{\mp }\mathcal{Q}_{\mp }(\epsilon ^{+}%
\mathcal{Q}_{+}+\epsilon ^{-}\mathcal{Q}_{-})T,  \nonumber \\
\delta _{a}(\delta _{\epsilon }\mathcal{A}_{\pm }) &=&i\epsilon ^{\mp
}\delta _{a}T=i\epsilon ^{\mp }(a^{+}\partial _{+}+a^{-}\partial _{-})T.
\label{120}
\end{eqnarray}
In the first line we have used of $\delta _{\epsilon }T=-\overline{\epsilon }%
\mathcal{Q}T,$ which has a minus sign contrary to the naive expectation.
Hence using the algebra of $\mathcal{Q}$'s we find 
\begin{eqnarray}
\lbrack \delta _{\epsilon },\delta _{a}]\mathcal{A}_{\pm } &=&a^{\mp
}(\epsilon ^{+}\mathcal{Q}_{\mp }\mathcal{Q}_{+}+\epsilon ^{-}\mathcal{Q}%
_{\mp }\mathcal{Q}_{-})T-\epsilon ^{\mp }(a^{+}\mathcal{Q}_{+}^{2}+a^{-}%
\mathcal{Q}_{-}^{2})T  \nonumber \\
&=&\mathcal{Q}_{\pm }(a^{+}\epsilon ^{-}\mathcal{Q}_{+}+a^{-}\epsilon ^{+}%
\mathcal{Q}_{-})T\equiv \mathcal{Q}_{\pm }\Lambda (a,\epsilon ),  \label{121}
\end{eqnarray}
which is a gauge transformation on $\mathcal{A}_{\pm }$ with the parameter 
\begin{eqnarray}
\Lambda (a,\epsilon ) &\equiv &(a^{+}\epsilon ^{-}\mathcal{Q}%
_{+}+a^{-}\epsilon ^{+}\mathcal{Q}_{-})T  \nonumber \\
&=&a^{+}\epsilon ^{-}O_{+}-a^{-}\epsilon ^{+}O_{-}.  \label{122}
\end{eqnarray}
Thus in this case the gauge parameter is a linear combination of the two
gauge invariants $O_{\pm }$. Accordingly, on $S$ we have 
\begin{equation}
\delta _{\epsilon }(\delta _{a}S)=-iS(a^{+}\delta _{\epsilon
}A_{+}+a^{-}\delta _{\epsilon }A_{-})+(a^{+}\nabla _{+}+a^{-}\nabla
_{-})\delta _{\epsilon }S.  \label{123}
\end{equation}
Using the definition of $A_{\pm }$ (eq.(\ref{109})) we find 
\begin{equation}
\delta _{\epsilon }A_{\pm }=-\epsilon ^{\mp }\mathcal{Q}_{\pm }T,
\label{124}
\end{equation}
upon using which the last expression becomes 
\begin{equation}
\delta _{\epsilon }(\delta _{a}S)=iS(a^{+}\epsilon ^{-}\mathcal{Q}%
_{+}T+a^{-}\epsilon ^{+}\mathcal{Q}_{-}T)+i(a^{+}P_{+}+a^{-}P_{-})(\epsilon
^{+}Q_{+}+\epsilon ^{-}Q_{-})S.  \label{125}
\end{equation}
Similarly, we find 
\begin{equation}
\delta _{a}(\delta _{\epsilon }S)=-iS(\epsilon ^{+}a^{-}\mathcal{Q}%
_{-}T+\epsilon ^{-}a^{+}\mathcal{Q}_{+}T)+i(\epsilon ^{+}Q_{+}+\epsilon
^{-}Q_{-})(a^{+}P_{+}+a^{-}P_{-})S.  \label{126}
\end{equation}
Subtracting the last two expressions gives the expected result 
\begin{eqnarray}
\lbrack \delta _{\epsilon },\delta _{a}]S &=&2iS(a^{+}\epsilon ^{-}\mathcal{Q%
}_{+}T+a^{-}\epsilon ^{+}\mathcal{Q}_{-}T)+i[a^{+}P_{+}+a^{-}P_{-},\epsilon
^{+}Q_{+}+\epsilon ^{-}Q_{-}]S  \nonumber \\
&=&2iS(a^{+}\epsilon ^{-}\mathcal{Q}_{+}T+a^{-}\epsilon ^{+}\mathcal{Q}%
_{-}T)-iS(a^{+}\epsilon ^{-}\mathcal{Q}_{+}T+a^{-}\epsilon ^{+}\mathcal{Q}%
_{-}T)  \nonumber \\
&=&i(a^{+}\epsilon ^{-}\mathcal{Q}_{+}T+a^{-}\epsilon ^{+}\mathcal{Q}%
_{-}T)S=i\Lambda (a,\epsilon )S,  \label{127}
\end{eqnarray}
where we have used the previous expressions for the commutators of $P_{\pm }$
with $Q_{\pm }$. This concludes our proof of the closure of the algebra of
supertranslations and super gauge symmetries in a GSFT.

The above computations reveal a generic structure: when evaluating a generic
commutator of two \textit{non-gauge} transformations like $[\delta
_{G_{1}},\delta _{G_{2}}]$ on $S$, we encounter two types of terms: one is
due to the variation of $S$ only, which has the form $-[G_{1},G_{2}]S$,
while the other is due to the variation of $\mathcal{A}_{\alpha }$ having
the form $2[G_{1},G_{2}]S$. The two terms have intriguingly similar forms so
that they add up to $[G_{1},G_{2}]S$; i.e. minus the expression would be
obtained if we had considered the variation of $S$ only. In other words 
\begin{equation}
\lbrack \delta _{G_{1}}\delta _{G_{2}}]=\delta _{[G_{1},G_{2}]}.  \label{128}
\end{equation}
Now, since $[G_{1},G_{2}]$ is a gauge invariant quantity, the parameters of
the transformations on the RHS become also gauge invariant.

Another consequence of these calculations is that the\thinspace gauge
invariant operators $T,O_{\pm },F$ appearing on the RHS\ of the algebra of $%
P_{\pm },Q_{\pm }$ may be interpreted as a basis of expansion for the
parameter $\Lambda (x,\theta )$ of the gauge transformations and, hence, as
the generators of these transformations.. This is because, as we have seen,
the parameters of the transformations on the RHS are gauge invariant
quantities proportional these gauge invariant operators.

\smallskip

\smallskip

\section{\protect\smallskip \textbf{\ Conclusion}}

In this paper, using the notion of a gauge superfield $\mathcal{A}_{\alpha }$
as a gauge connection on superspace, we have constructed a generalized class
of SUSY algebras, which besides including the usual (super)translation
generators $P_{\pm }$, $Q_{\pm }$, they involve the new generators $%
T,F,O_{\pm }$ corresponding to the generators of the $U(1)$ gauge symmetry
associated to $\mathcal{A}_{\alpha }$. We found that this algebra is \textit{%
gauge invariant, }but\textit{\ }in general \textit{not closed} in the sense
of a super Lie algebra, if we treat $\mathcal{A}_{\alpha }$ as some \textit{%
fixed} superfield. Nevertheless, it is closed\smallskip for very specific
configurations of $\mathcal{A}_{\alpha }$ (or more precisely, its field
strength $T$). Two examples within this class are the centrally extended
supersymmetry (CE-SUSY) and the non-commutative supersymmetry (NC-SUSY)
algebras. These are indeed the only rotationally symmetric configurations
for which the algebra also closes with the $SO(2)$ rotation generator $J$.
There are, however, other non-rotationally symmetric configurations of the
gauge superfield on which also the algebra closes (but of course without the 
$SO(2)$ generator).

\smallskip

We showed that the generalized SUSY algebra is realized in a gauge
superfield theory (GSFT) which, in addition to a global supersymmetry, it
also possesses a local superspace gauge symmetry. In this case, the
generalized supertranslation symmetries are realized as a result of knitting
these (local and global) symmetries of the theory in a particular way. As
transformations changing \textit{both} the gauge and matter superfields $(%
\mathcal{A}_{\alpha },S)$, the supertranslations and super gauge
transformations were found to form a closed algebra. We noted that, in this
interpretation, the generalized supersymmetry is realized by a \textit{%
non-linear} realization of the corresponding transformations $\delta
_{a},\delta _{\epsilon },\delta _{\Lambda }$ on $(\mathcal{A}_{\alpha },S)$.
The algebra is then as follows 
\begin{eqnarray}
\lbrack \delta _{\epsilon },\delta _{\epsilon ^{\prime }}] &=&\delta
_{a}+\delta _{\Lambda },\qquad [\delta _{a},\delta _{a^{\prime }}]=\delta
_{\Lambda },\qquad [\delta _{\epsilon },\delta _{a}]=\delta _{\Lambda }, 
\nonumber \\
\lbrack \delta _{\Lambda },\delta _{\epsilon }] &=&0,\qquad [\delta
_{\Lambda },\delta _{a}]=0,\qquad [\delta _{\Lambda },\delta _{\Lambda
^{\prime }}]=0.\qquad   \label{129}
\end{eqnarray}
The last commutation relation accounts for the fact that the gauge symmetry
under consideration is an abelian one (in the superspace sense it is a $U(1)$
gauge symmetry). It would be interesting to find the non-abelian
generalization of the above constructions. As might be expected, the above
algebra generalizes the ordinary supertranslation algebra by replacing any
vanishing commutator in the ordinary case by a gauge transformation. We
computed the dependences of the transformation parameters on the RHS of
these equations to the parameters on their LHS.

\smallskip

\smallskip

\textbf{\large Acknowledgment}

The author would like to thank M.H. Sarmadi for useful discussions on this
work and H. Fakhri for his concern and continuous encouragements.

\section{Appendix}

\renewcommand{\theequation}{A.\arabic{equation}}\setcounter{equation}{0}

In this appendix we collect some definitions, conventions and properties
which are required in the course of this paper.

The complex coordinates $x^{\pm }$ and their derivatives are defined in
terms of the real coordinates $(x^{1},x^{2})$ as 
\begin{eqnarray}
x^{\pm } &\equiv &\frac{1}{2}(x^{1}\pm ix^{2}),  \nonumber \\
\partial _{\pm } &\equiv &\partial _{1}\mp i\partial _{2}.  \label{130}
\end{eqnarray}
We use the complex spinors $\theta $ with the components $\theta ^{\pm }$
related by conjugation 
\begin{equation}
(\theta ^{+})^{*}=\theta ^{-}.  \label{131}
\end{equation}
The inner product of two spinors $\epsilon $ and $\theta $ is defined in
terms of their components as follows 
\begin{equation}
\overline{\epsilon }\theta \equiv \epsilon ^{+}\theta _{+}+\epsilon
^{-}\theta _{-},  \label{132}
\end{equation}
The spinorial indices are raised with $\varepsilon ^{\alpha \beta }$ and
lowered with $\varepsilon _{\alpha \beta }$, which means that 
\begin{equation}
\theta _{+}=\theta ^{-},\qquad \theta _{-}=-\theta ^{+}.  \label{133}
\end{equation}
We can check that the above inner product is always real; i.e. 
\begin{equation}
(\overline{\epsilon }\theta )^{*}=\overline{\epsilon }\theta  \label{134}
\end{equation}
Under a $SO(2)$ rotation with an angle $\alpha $, the spinor and the vector
components with higher indices are transformed as 
\begin{equation}
\theta ^{\pm }\rightarrow e^{\pm i\alpha /2}\theta ^{\pm },\qquad x^{\pm
}\rightarrow e^{\pm i\alpha }x^{\pm },  \label{135}
\end{equation}
while those with lower indices are changed by the inverses of these
transformations.

\smallskip

Supertranslations are defined as 
\begin{equation}
\delta \theta ^{\pm }=\epsilon ^{\pm },\qquad \delta x^{\pm }=i\epsilon
^{\pm }\theta ^{\pm }.  \label{136}
\end{equation}
The corresponding ordinary translation and supertranslation generators are 
\begin{equation}
\mathcal{P}_{\pm }=-i\partial _{\pm },\qquad \mathcal{Q}_{\pm }=\frac{%
\partial }{\partial \theta ^{\pm }}+i\theta ^{\pm }\partial _{\pm }.
\label{137}
\end{equation}
The ordinary supertranslation algebra is then given by 
\begin{eqnarray}
\mathcal{Q}_{\pm }^{2} &=&-\mathcal{P}_{\pm },\qquad \{\mathcal{Q}_{+},%
\mathcal{Q}_{-}\}=0,  \nonumber \\
\lbrack \mathcal{P}_{+},\mathcal{P}_{-}] &=&0,\qquad [\mathcal{P}_{\pm },%
\mathcal{Q}_{\pm }]=0.  \label{138}
\end{eqnarray}
The conjugation properties of these operators are 
\begin{equation}
(\mathcal{P}_{\pm })^{\dagger }=\mathcal{P}_{\mp },\qquad (\mathcal{Q}_{\pm
})^{\dagger }=\mathcal{Q}_{\mp },  \label{139}
\end{equation}
while those for their action on a complex (grassmann even) superfield $S$
are 
\begin{equation}
(\mathcal{P}_{\pm }S)^{*}=-\mathcal{P}_{\mp }S^{*},\qquad (\mathcal{Q}_{\pm
}S)^{*}=\mathcal{-Q}_{\mp }S^{*}.  \label{140}
\end{equation}
For a grassmann odd superfield the sign on the RHS of the second equation is
reversed.

\end{document}